\documentclass[doublecol]{epl2}
\usepackage{amssymb,amsmath,latexsym}
\usepackage{epsfig}
\def\Fbox#1{\vskip1ex\hbox to 8.5cm{\hfil\fboxsep0.3cm\fbox{%
  \parbox{8.0cm}{#1}}\hfil}\vskip1ex\noindent}  

\def\Fbox#1{\vskip1ex\hbox to 8.5cm{\hfil\fboxsep0.3cm\fbox{%
  \parbox{8.0cm}{#1}}\hfil}\vskip1ex\noindent}  

\let \= \equiv \let\*\cdot \let\~\widetilde \let\-\overline

\begin{document}
\title{The nature of the $\beta$-peak in the loss modulus of amorphous solids}
\author{Yossi Cohen$^1$, Smarajit Karmakar$^{2}$, Itamar Procaccia$^1$ and Konrad Samwer$^3$}
\institute{$^1$Department of Chemical Physics, The Weizmann
 Institute of Science, Rehovot 76100, Israel. \\ $^2$ Dept of Physics, Universita di Roma La Sapienza, Piazale Aldo Moro 2, Roma.\\ $^3$ Dept of Physics, University of G\"ottingen, Germany.}
\abstract{
Glass formers exhibit, upon an oscillatory excitation, a response function whose imaginary and real parts are known as the loss and storage moduli respectively. The loss modulus typically peaks at a frequency known as the $\alpha$ frequency which is associated with the main relaxation mechanism of the super-cooled liquid. In addition, the loss modulus is decorated by a smaller peak, shoulder or wing which is referred to as the $\beta$-peak. The physical origin of this secondary peak had been debated for decades, with proposed mechanisms ranging from highly localized relaxations to entirely cooperative ones. Using numerical simulations we bring an end to the debate, exposing a clear and unique cooperative mechanism for the said $\beta$-peak which is distinct from that of the $\alpha$-peak.}

\maketitle

\begin{figure}
\begin{center}
\includegraphics[scale = 0.450]{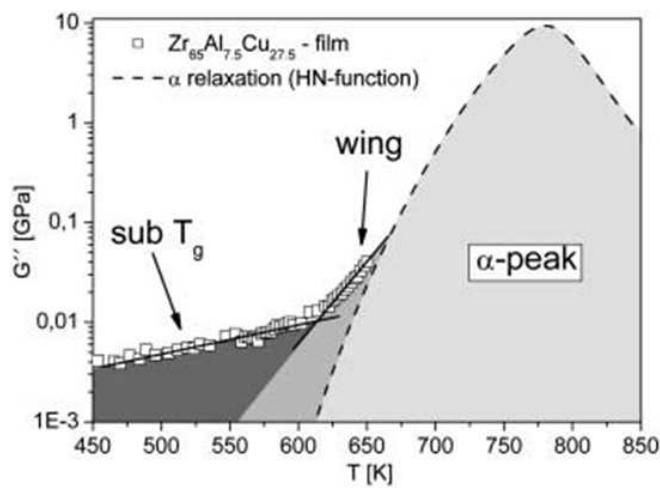}
\caption{Color online; a typical loss modulus as a function of temperature. Shown are experimental results for
a film of the metallic glass
Zr$_{65}$Al$_{7.5}$Cu$_{27.5}$ which was forced at a fixed frequency $\omega=5440Hz$. Note the well defined $\alpha$ peak and
the shoulder, or wing, which appears at lower temperatures and is referred to as the $\beta$ peak.\cite{04RSL}.Havriliak-Negami (HN) curve
is plotted using parameter $\alpha = 1$ and $\gamma = 0.6$. Please see Ref.\cite{04RSL} for the form of the HN function and other details.}
\label{lossmodulus}
\end{center}
\end{figure}

{\bf Introduction}: In Fig. \ref{lossmodulus} one observes a typical loss modulus, in this case of a film of the metallic glass
Zr$_{65}$Al$_{7.5}$Cu$_{27.5}$ which was forced at a fixed frequency $\omega=5440Hz$ while the temperature was changed from 450K
to 850K \cite{04RSL}.
The main peak represents the typical $\alpha$ relaxation which is the slowest mode of relaxation which is typical to glass formers.
When the temperature was at a value such that this relaxation frequency was at resonance with the fixed forcing frequency, the loss
of energy registered a maximum. Upon reducing the temperature one sees a wing which becomes obvious at about 650K, this wing is
associated with a typical
frequency that is higher than the $\alpha$ frequency. This is referred to as a secondary relaxation or the $\beta$ wing, and in
different glassy materials it
appears more or less prominent, sometime as a separate peak, and sometime as a shoulder or a wing \cite{00Loi}. Note at this point
that the
literature is full of other phenomena in glassy relaxation that are referred to as $\beta$ \cite{70JG,09Cav,11BB,06D,99G,04NP}.
In this Letter we only deal with the one
presented here as in Fig. \ref{lossmodulus}. This is also referred to as the Johari-Goldstein
process or the slow $\beta$-process~\cite{70JG}.

The existence of the $\beta$-peak was known for a long time, but its physical origin remained debated. Past explanations can be
roughly grouped into two main
groups, those involving an extension of the $\alpha$-relaxation process and those proposing an independent relaxation mechanism.
In the first group one can quote for example Dixon et al. \cite{90DNWC}, whereas
the second group is exemplified by \cite{00SBLL}. There seems to be no consensus even on the question whether the $\beta$ process
is local or cooperative. The aim of this Letter is to disperse the fog that had accumulated over the physics of the $\beta$-peak
and to reach a unique interpretation of the relevant physics. To this aim we use numerical simulations as described next. Of
course, numerical simulations are
limited in their ability to probe long times, and we are therefore constrained to examine temperatures
for which the increase in glassy relaxation times is only 3 to 4 orders of magnitude compared to twelve to fourteen orders in
experiments. Nevertheless, we do not expect a fundamental change in physics as the
system cools down towards longer and longer relaxation times. This is the basic assumption on which our
conclusions rely.

{\bf Numerical simulations}: As a first step we have reproduced the phenomenology of the $\beta$-peak
on the machine. To this aim we employed as an example one of the standard models of glass formers, i.e. a binary mixture of
point particles interacting via Lennard-Jones potential with three different characteristic interaction lengths
$\sigma_{ss}=1$,$\sigma_{\ell\ell}=1.4$  and
$\sigma_{s\ell}=1.18$. Details of the potentials can be found for example in Ref. \cite{11KLPZ}. Of course the phenomenon
discussed appears in other models in much the same way, as we have checked by running similar simulations on other models.
In the present simulations the unit of length is $\sigma_{ss}$, temperature is measured in units such that the Boltzmann's
constant equals unity, and time in units of inter-particle distance over the mean velocity. Two dimensional samples, each
containing $4900$ particles, were firstly quenched slowly (reducing the temperature by $5\times 10^{-4}$ in every step) to
an inherent state configuration at $T=0$ using a gradient energy method. We employ periodic boundary conditions keeping the
volume of the box fixed. In a second step the samples were raised to a chosen temperature $T$ and were
subjected to an oscillatory shear strain $\gamma=\gamma_0 \sin(\omega t)$ according to the affine transformation
\begin{equation}
x_i\rightarrow x_i + (\dot{\gamma}dt) y_i, \ y_i\rightarrow y_i \ .
\end{equation}
in addition we impose Lees-Edwards boundary conditions. The maximum affine strain is $\gamma_0=0.005$, and $x_i, y_i$ are
the position of the $i$'th particle. The Gear predictor-corrector algorithm was used to integrate the sllod equations of
motion \cite{AT91}.

The stress in the system was measured using 50 different realizations, and the mean shear stress was fitted to an oscillatory
function $\tau_{xy}=\tau_0 \sin (\omega t + \delta)$. The $\delta$ corresponds to the phase shift between the strain and
the stress. Now, we can calculate the dynamic shear modulus,
\begin{equation}
G=\frac{\tau_0}{\gamma_0}e^{i\delta}
\end{equation}
where $G=G^{'} + iG^{''}$ is a complex  number that depends on the temperature and the shearing frequency. $G^{'}$ and $G^{''}$
are known as the storage and the loss moduli.

\begin{figure}
\begin{center}
\includegraphics[scale = 0.550]{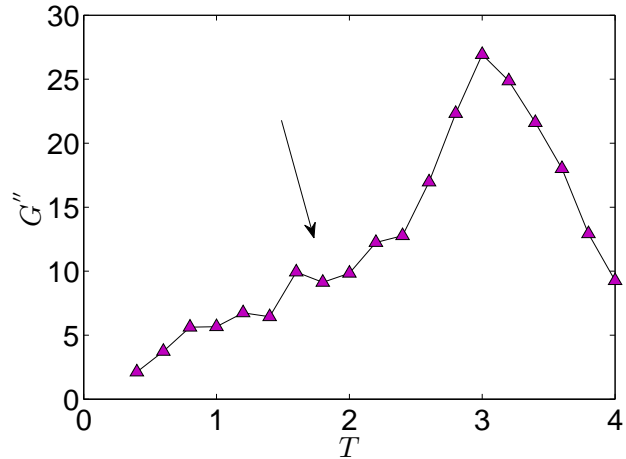}
\caption{Color online; the loss modulus $G^{''}$ as a function of temperature from numerical simulations ($\omega=0.01$).
Observe the clearly defined $\alpha$ peak and the distinct $\beta$ wing. }
\label{losssim}
\end{center}
\end{figure}

To produce data which are analogous to the experimental ones in Fig. \ref{lossmodulus} we studied $G^{''}$ when the system
was subject to such a simple oscillatory shear with a fixed frequency $\Omega$
for a range of temperatures. A typical result is shown in Fig. \ref{losssim} which pertains to $\omega=0.01$ and $T$ in the
range 0.5$\le T\le$6. In a satisfactory agreement with the experimental picture we observe a clear $\alpha$-peak decorated
with a distinct $\beta$-wing in the range of temperatures 1.0$\le T \le$2.0.
Similar simulations done at a fixed temperature with varying frequencies resulted in the same evidence
for the existence of the $\beta$-process as seen in experiments. We thus feel encouraged to proceed
to study the nature of this process.

\begin{figure}
\begin{center}
\includegraphics[scale=0.55]{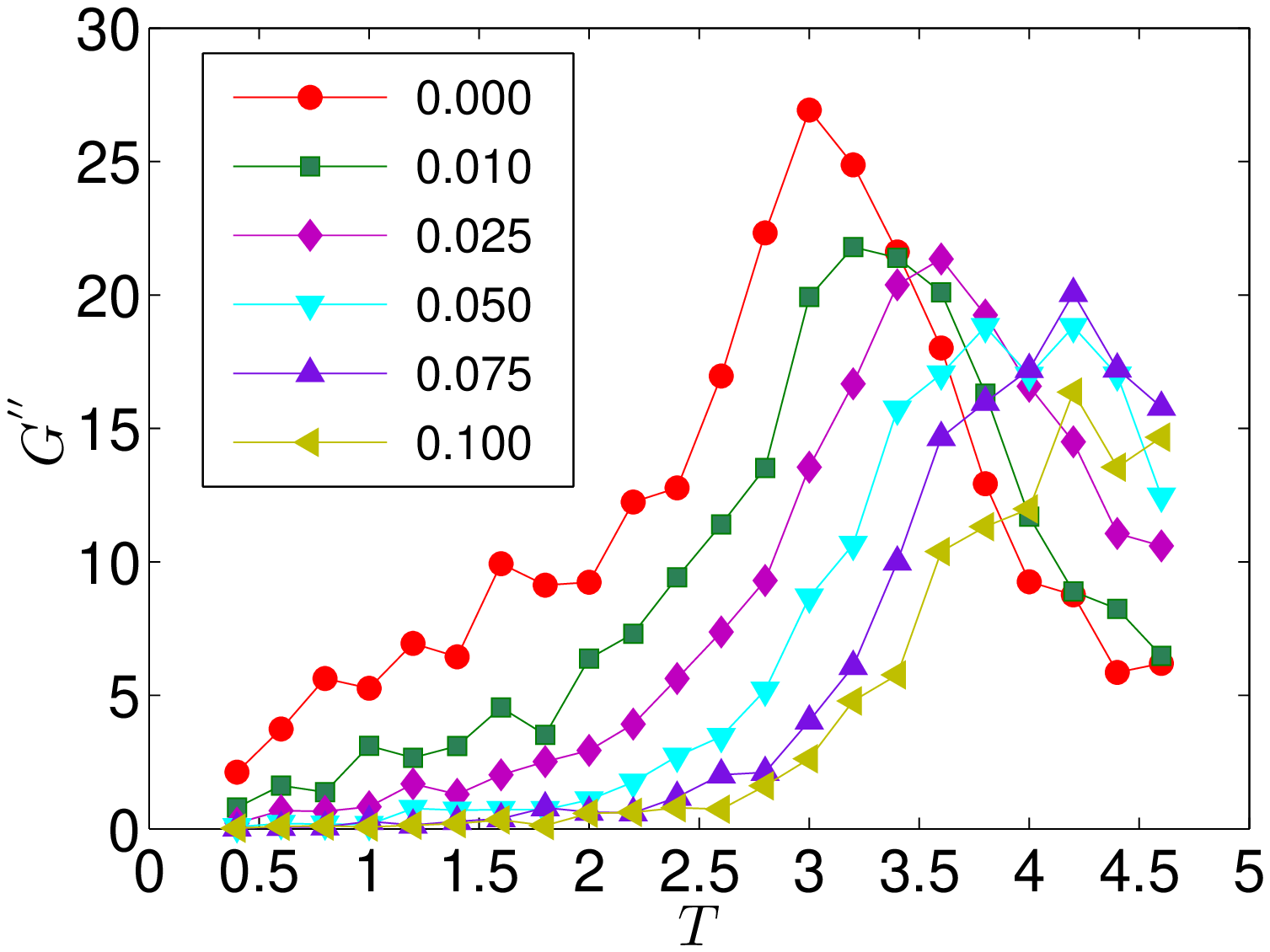}\label{ic1}
\includegraphics[scale=0.54]{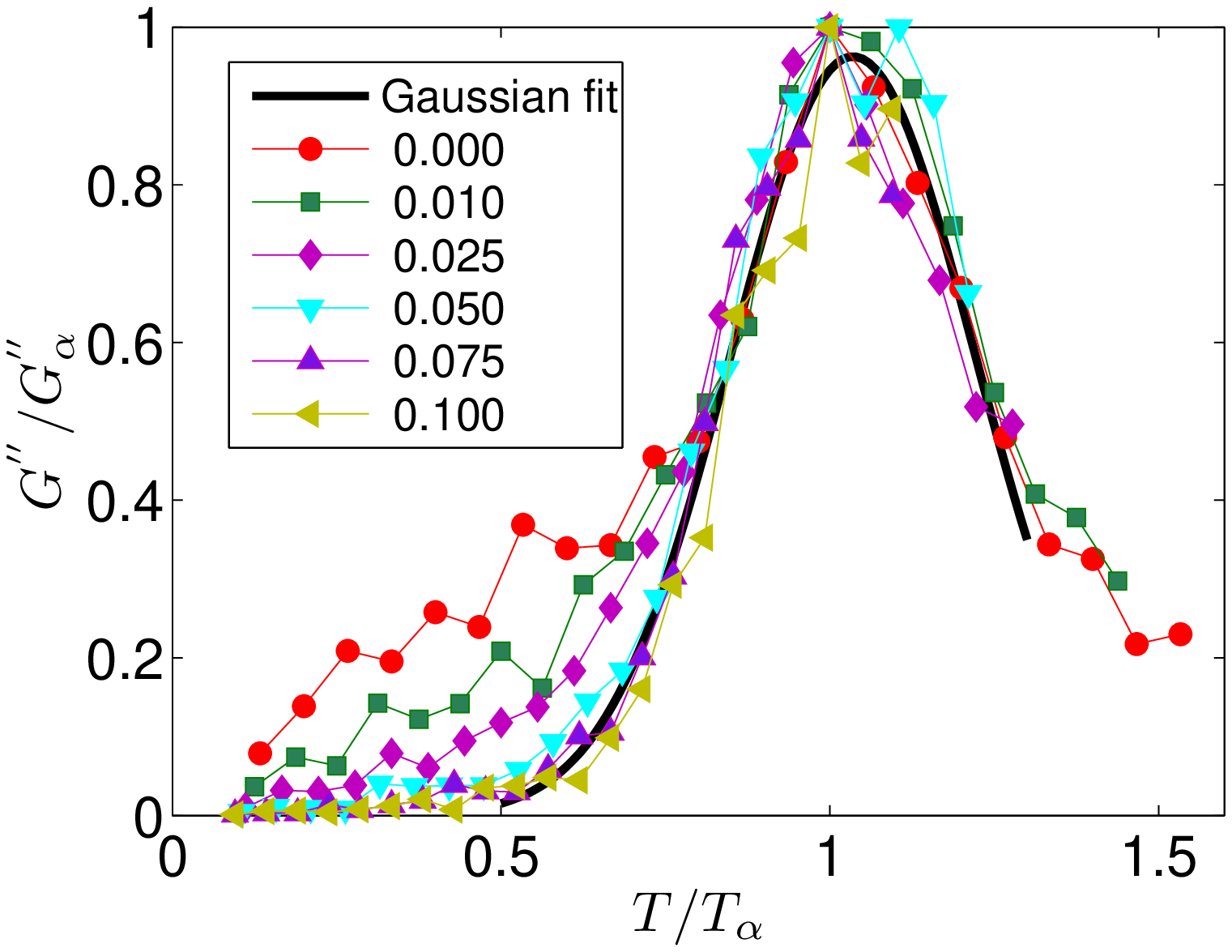}\label{ic2}
\caption{Color online; Upper panel: the loss modulus as shown in the previous figure, and the loss modulus when a small fraction of
particles get pinned. The consequence of varying concentrations of pinned particles are shown. Note that it is sufficient to pin
more than 2.5\% of the particles to eliminate the $\beta$ wing altogether. Lower panel: collapse of the data shown in the upper
panel, obtained by rescaling the temperature with a characteristic $G^{''}_\alpha$ and $T_\alpha$, the value and the position of
the $\alpha$ peak which depends on the concentration of pinned particles. We see that the $\beta$ peak is eliminated with all
the concentrations larger than 2.5\%. The grey curve is a Gaussian fit to the remaining $\alpha$ process.}
\label{pinned}
\end{center}
\end{figure}

{\bf The $\beta$-process, local or cooperative?}: presently we can use the power of numerical simulations
to answer the first crucial question, is the $\beta$-process local or cooperative? To this aim we have selected randomly a small
fraction of the particles in our system, and {\bf pinned} them to the coordinate frame that moves with the affine transformation,
not allowing them to participate in the relaxation dynamics. We expect that such a procedure should certainly affect the
$\alpha$-peak which is known to be cooperative \cite{05BBB, 12HKPZ}. As for the $\beta$-wing, if it is associated with a
local process, a small fraction of pinned particles should not make a big difference. It turns out that a small fraction of
pinned particles, say larger than 2.5\%, is sufficient to suppress the $\beta$-wing altogether. In Fig. \ref{pinned} we show
the results of this exercise with varying percentage of pinned particles, from $0\%$ to 10\%. While the $\alpha$-peak is affected
as expected, slightly suppressed and pushed to higher
temperatures (the decay process is pushed to lower frequencies)\cite{06D}, the $\beta$-wing is suppressed altogether when the
concentration of the pinned particles exceeds 2.5\%. We can thus conclude that the $\beta$-process must be cooperative rather
than local. This immediately excludes the first group of mechanisms discuss above.

\begin{figure}
\begin{center}
\includegraphics[scale = 0.50]{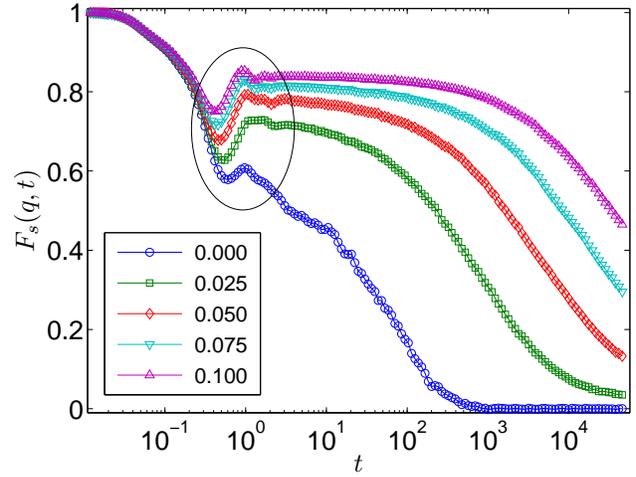}
\caption{Color online. The self part of the intermediate scattering function for $T=3$ for several pinning densities in a 2D
system. Note the change of the shape of the function in the encircled area.}
\label{isf}
\end{center}
\end{figure}

{\bf Which cooperative mechanism then?}:  to unfathom the nature of the cooperative $\beta$ process,
we examined carefully the dynamics in our system on time scales of the order of the cage dynamics, i.e.
for time scales of the order of unity in our re-scaled units.
In Fig. \ref{isf} we show the self part of the intermediate scattering function $F_s(q,t)$ for a fixed
temperature and various pinning density. $F_s(q,t)$ is defined as
\begin{equation}
F_s(q,t) = \frac{1}{N'}\left \langle \sum_{j=1}^{N'} e^{\imath \vec{q}\cdot[\vec{r}_j(t) - \vec{r}_j(0)]} \right \rangle,
\end{equation}
where $N'$ is the number of particles which are mobile, $\langle \ldots \rangle$ represent thermal average and $\vec{r}_j(t)$
is the position vector of particle $j$ at time $t$.
$\vec{q}$ is chosen uniformly on the surface of a sphere of
radius $q = |\vec{q}| = 2\pi/r_0$, where $r_0 = 2^{1/6}$ is the position of the minimum in the potential function as a function of distance.
This value roughly corresponds to the first peak position in the static structure factor.
We took $128$ different directions for the averaging of $F_s(q,t)$. We observe a strong slowing down in the dynamics as we increase
the pinning density: the effective $\tau_{\alpha}$
becomes larger. But we also observe a hump in the curve, on time-scales corresponding to the $\beta$ process, which almost
disappear when the concentration of the pinned particles vanishes. This hump can be explained by a recoil motion of a free
particle when it collides with a pinned particle. Evidently this phenomenon is less pronounced when the collision occurs
with two free particles. In this case, one particle can transfer momentum to its neighbor etc, a process that evolves into
a collective motion of number of particles due to the collisions. This cooperative motion is suppressed by the pinned particles.

\begin{figure*}
\includegraphics[scale = 0.49]{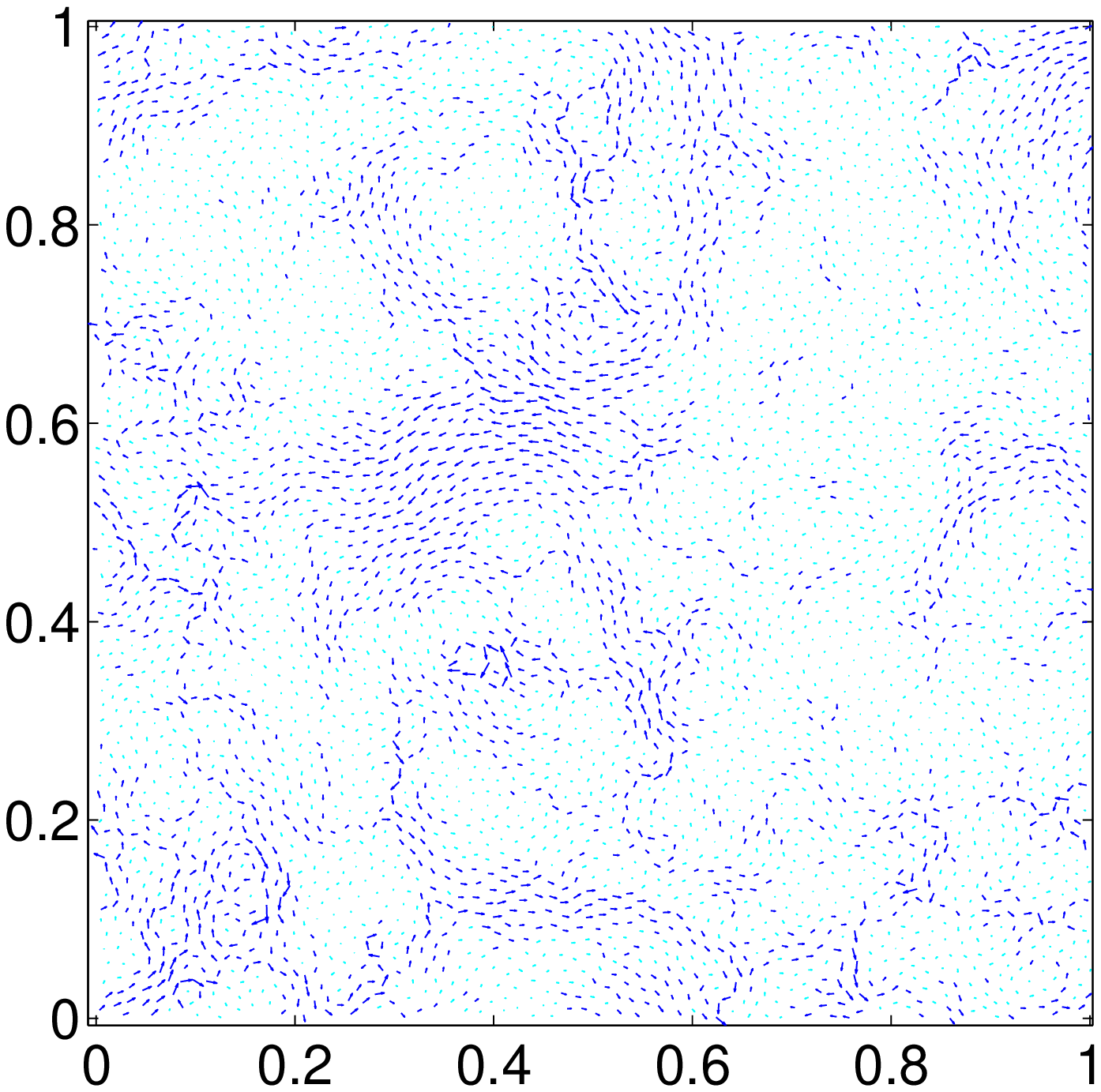}
\hskip -0.7cm
\includegraphics[scale = 0.49]{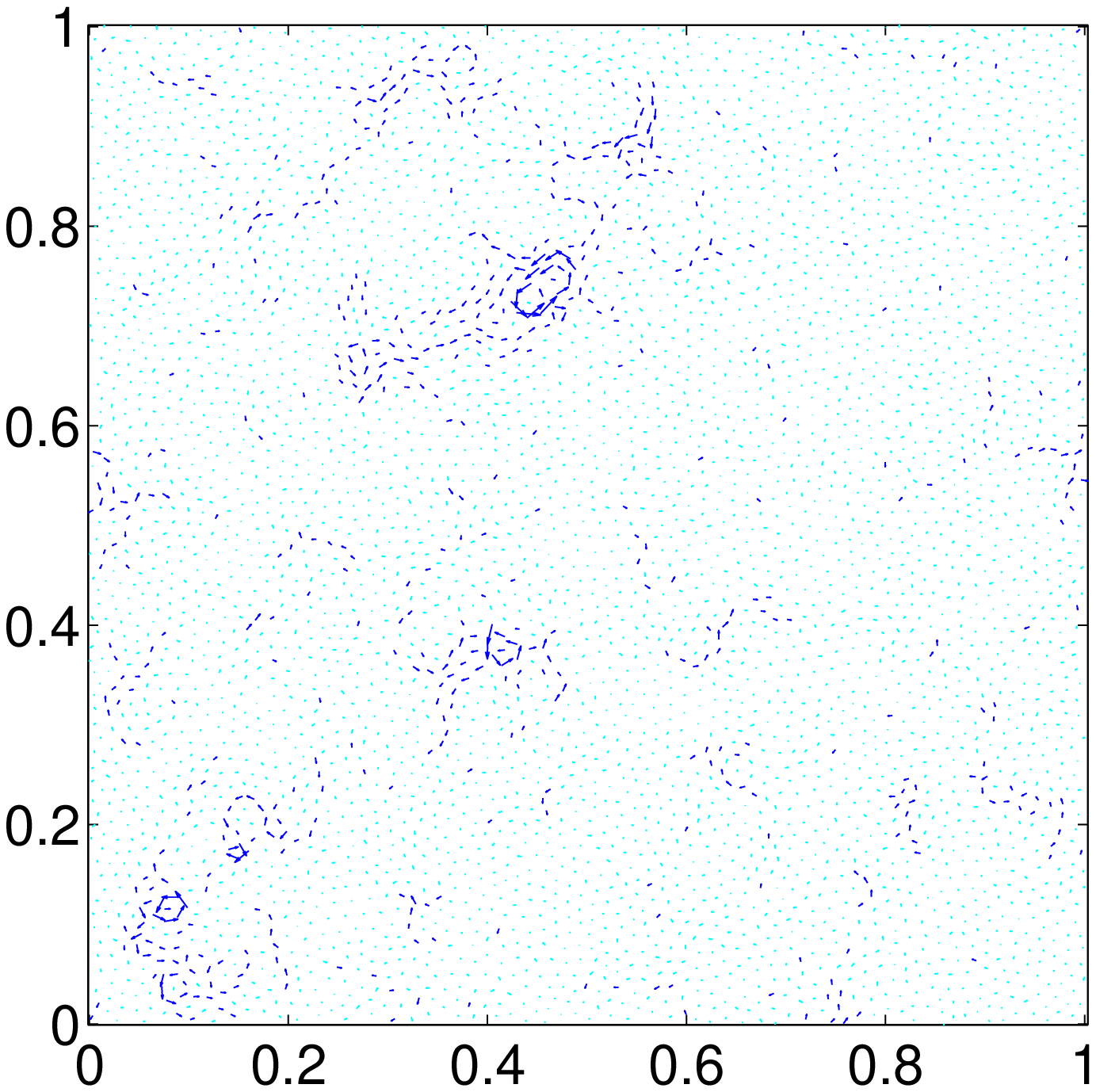}
\caption{Color online: Left Panel:  a graphic representation of the cooperative motion that is associated with the $\beta$ wing.
Note the chains of particles that have moved coherently during a time span of four time units. The particles that move more than
$40\%$ of the typical inter-particle distance are marked in dark blue. Right Panel: similar graphic representation of the
suppression of the majority of the cooperative motion that is responsible for the $\beta$ wing by the addition of 2.5\% pinned
particles. In contrast to the previous figure, here one needs to look at cumulated motions for 15 time units to see the remnant
correlated motion.}
\label{free}
\end{figure*}

In left panel of Fig. \ref{free} we show the cumulative displacement of all the particles after four time units (i.e. at $t=4$
with respect to their position at $t=0$).
We observe long chains of correlated motion of particles all occurring at thermal equilibrium which is maintained only by the
influence of random forces. We emphasize that no external forces were applied. The darker arrows are to indicate the particles
that move more than $40\%$ of the typical inter-particle distance.
In comparison, with 2.5\% of pinned particles, there is hardly any such correlated motions on the scale of 4 units of time.
One has to increase the time window to about 15 units of time, cf. the right panel of Fig. \ref{free}, in order to observe
the remnant weak cooperative motion.


In summary, we have shown here that the $\beta$ wing measured in loss moduli has to do with a cooperative motion which is
however associated with very fast time scales, of the order of the cage time. The cooperative motion consists of chains
of particles that move cooperatively when the system is forced in an oscillatory fashion. Thus cooperative yes, but an extension
of the $\alpha$ process no. We trust that the clear cut numerical experiments discussed above should remove any
doubt as to the origin of the $\beta $ wing in the loss modulus.

{\bf Acknowledgments}: This work had been supported in part by the German Israeli Foundation, the Israel Science Foundation,
and the European Research Council under an advanced 'ideas' grant.

\end{document}